\documentclass[twocolumn,prd,nofootinbib,floats,floatfix,amsmath,amssymb,secnumarabic]{revtex4-1} %
\pdfoutput=1
\usepackage{amsmath}
\usepackage{amsfonts}
\usepackage{amssymb}
\usepackage{latexsym}
\usepackage{graphicx}
\usepackage[english]{babel}
\usepackage{multirow}
\usepackage{float}
\usepackage{url}
\usepackage{slashed}
\usepackage{verbatim}
\usepackage{ulem,fancyvrb}
\usepackage{xcolor}

\newcommand{\be}{\begin{equation}}
\newcommand{\ee}{\end{equation}}
\newcommand{\ba}{\begin{array}}
\newcommand{\ea}{\end{array}}
\newcommand{\bea}{\begin{eqnarray}}
\newcommand{\eea}{\end{eqnarray}}

\def\roughly#1{\mathrel{\raise.3ex\hbox
{$#1$\kern-.75em\lower1ex\hbox{$\sim$}}}}

\begin{document}

\draft
\title{Viability of detection by AMS of sudden features due to dark matter annihilation to positrons and electrons}

\author{Arjun Sharma}
\affiliation{Department of Physics, University of Chicago, Chicago Il, 60637}



\date{\today}

\begin{abstract}
The Fermi experiment has measured the cosmic ray electron+positron spectrum and positron fraction [$\Phi_{e^+}/(\Phi_{e^+ + e^-})]$, and PAMELA has measured the positron fraction with better precision. While the majority of cosmic ray electrons and positrons are of astrophysical origin, there may also be a contribution from dark matter annihilation in the galactic halo. The upcoming results of  the AMS experiment will show measurements of these quantities with far greater precision. One dark matter annihilation scenario is where two dark matter particles annihilate directly to  $e^+$ and  $e^-$ final states. In this article, we calculate the signature ``bumps" in these measurements assuming a given density profile (NFW profile).  If the dark matter annihilates to electrons and positrons with a cross section $\sigma v \sim\,  10^{-26}$ cm$^3$/s  or greater, this feature may be discernible by AMS. However, we demonstrate that such a prominent spectral feature is already ruled out by the relative smoothness of the positron + electron cosmic ray spectrum as measured by Fermi. Hence we conclude that such a feature is undetectable unless the mass is less than $\sim$40 GeV.  
\end{abstract}

\maketitle


\begin{figure*}[!]
\vspace{-.5cm}
\includegraphics[width=0.46\textwidth ]{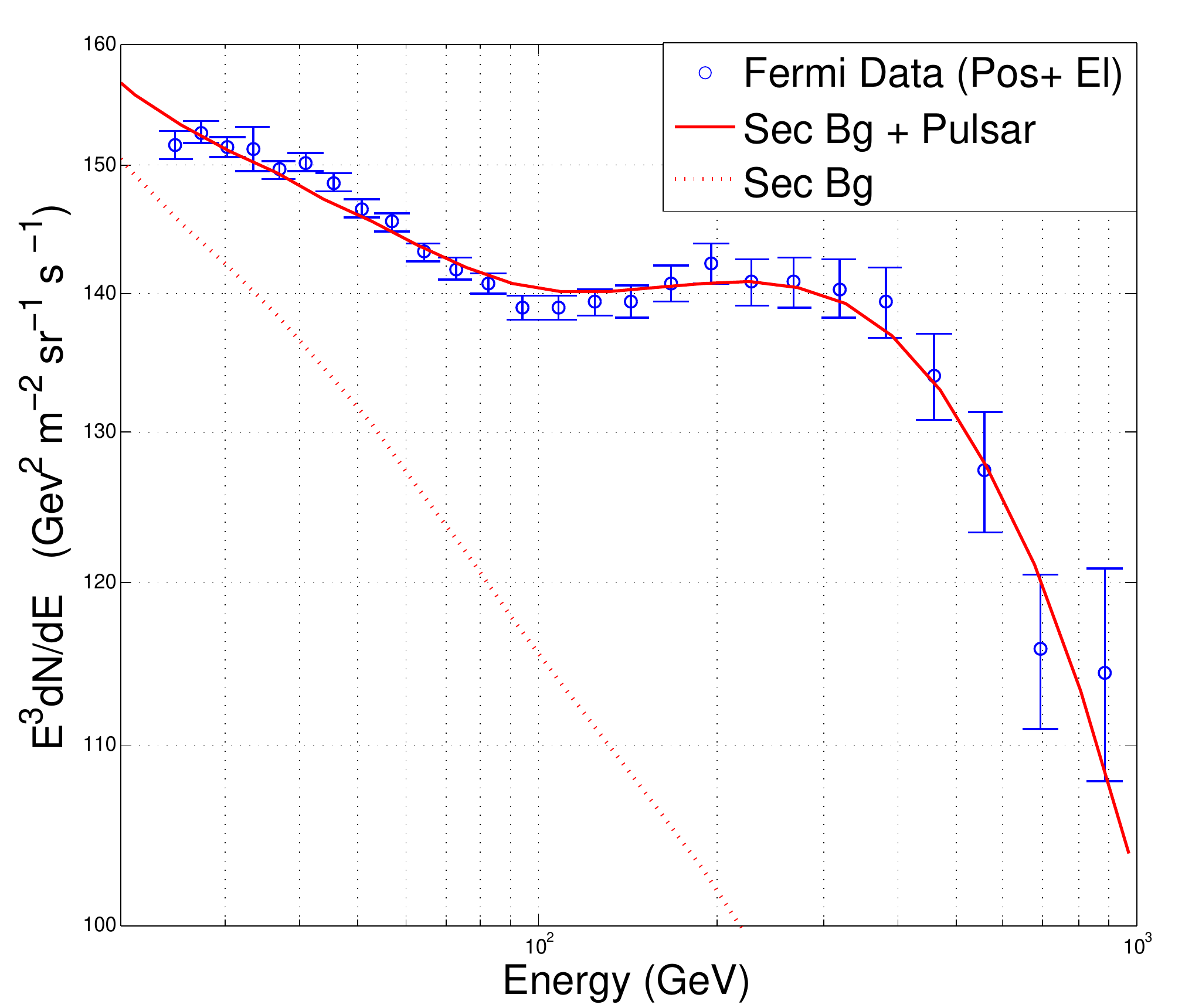}
\hspace{0.0cm}
\includegraphics[width=0.46\textwidth ]{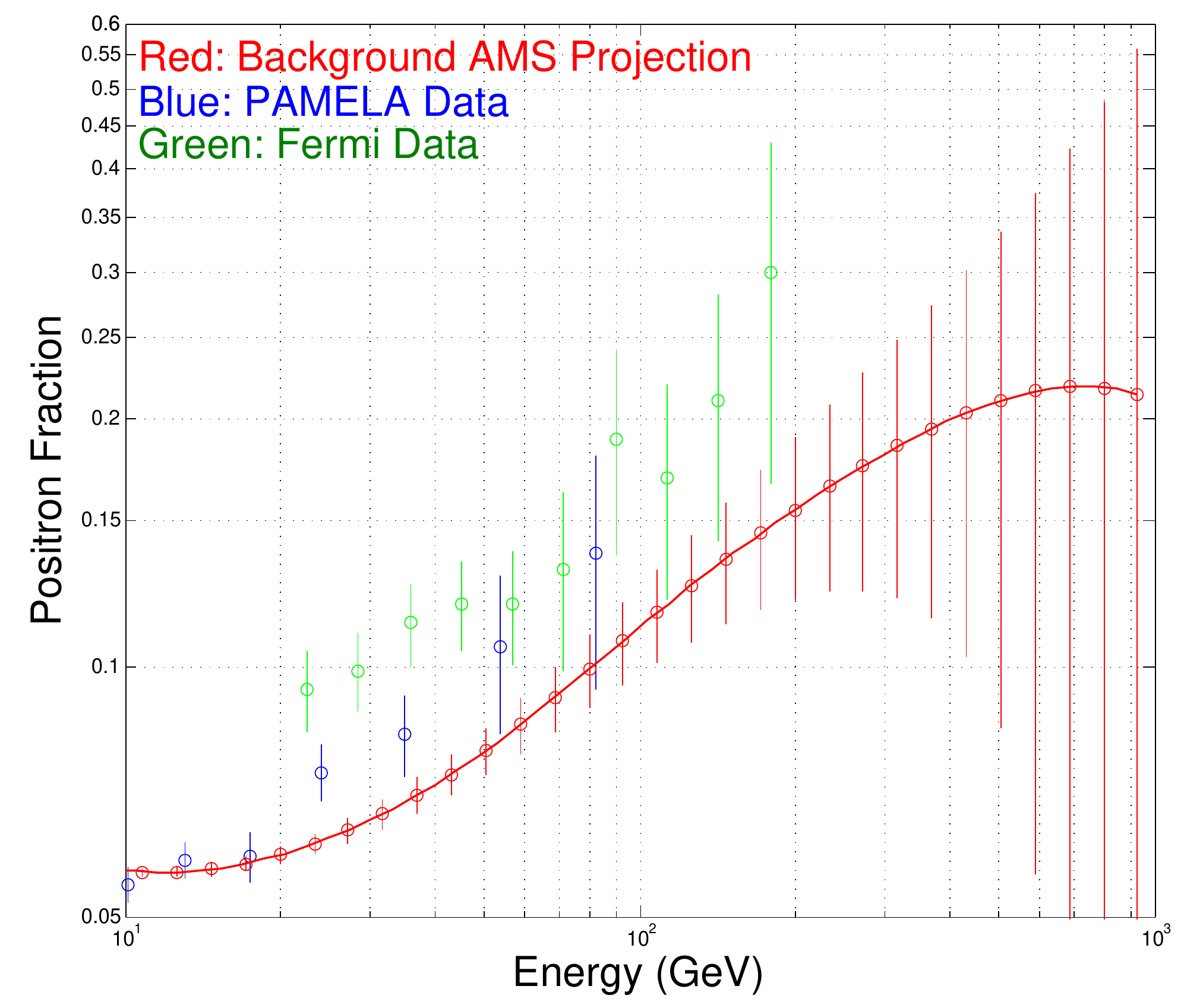} 
\vspace{0cm}
\caption{Left: Predicted positron + electron flux compared with Fermi data. Our model of background flux (dotted line) added to the flux from one pulsar (solid red) is a close match ( $\chi^2$= 34.7 for 26 degrees-of-freedom) of the Fermi data (blue error bars). Right: Corresponding positron fraction with projected error bars for AMS compared with PAMELA  and Fermi data. The pulsar flux is assumed to consist equally of electrons and positrons. See text for more details.}
\label{leptonflux}
\end{figure*}

\section{Introduction}
 From the measurements of the spectrum of cosmic ray  positrons and electrons experiments including Fermi Gamma Ray Space Telescope~\cite{FermiLAT:2011ab} and PAMELA ~\cite{Adriani:2008zr}, we know not only the individual spectra, but also the positron fraction [$\Phi_{e^+}/(\Phi_{e^+ + e^-})]$. These measurements of two related quantities, say $(x + y)$ and $x/(x+y)$, can be used to constrain the size of possible features in these spectra, and hence the origin of these cosmic rays. One possible source of these cosmic rays is dark matter annihilation. Of particular interest, is the case in which two dark matter particles annihilate to produce a positron and an electron. This leads to a signature ``bump" feature. 

The results of the Alpha Magnetic Spectrometer (AMS-02) experiment~\cite{AMS} are imminent. AMS will detect positrons and electrons in the cosmic ray spectrum between approximately 100 MeV and 1 TeV. With its much larger acceptance than PAMELA ($\sim$0.045 m$^2$sr vs. $\sim$0.002 m$^2$sr) ~\cite{Adriani:2008zr} and its high level of proton rejection, AMS is expected to measure the cosmic ray positron flux as well as the positron fraction  in far greater detail than was previously measured by PAMELA or Fermi. 

In this article, we consider the case in which the dark matter particles annihilate directly to electron-positron pairs, giving rise to an  edge-like feature in the cosmic ray positron spectrum at an energy equal to the mass of the annihilating WIMP~\cite{Baltz:2004ie, hooperbertone, hoopersilk}.  Because of its  increased precision, such features previously undetectable by PAMELA, might be detectable with AMS. We find that for dark matter masses greater than $\sim$40 GeV, cross sections required to get a detectable spectral feature in the positron fraction are already ruled out due to the smoothness of the positron + electron spectrum as measured by Fermi.



\section{Cosmic Ray Electrons and Positrons from Dark Matter Annihilation}

Positrons and electrons produced by galactic sources, as well  as due to dark matter annihilation, propagate through the galaxy under the influence of tangled magnetic fields. Here they lose energy through inverse Compton and synchrotron interactions ~\cite{baltzprop}. These effects can be modeled by the simple form of the diffusion-loss equation:
\begin{align}
\frac{\partial }{\partial t } \frac{dn}{d\epsilon}&=& \bigtriangledown\cdot\left[ K( \epsilon, \vec x  ) \vec\bigtriangledown \frac{\partial n}{\partial \epsilon} \right]+ \frac{\partial }{ \partial \epsilon }\left[ b(\epsilon, \vec x)  \frac{\partial n}{\partial \epsilon}\right] +  Q (\epsilon, \vec x)  
\end{align}
where $\epsilon =$ $E$/(1 GeV) parametrizes the energy of cosmic ray particles, $K$ is the diffusion constant (denoted $D_{xx}$ in GALPROP), $b$ is the energy loss rate, and Q the source term, i.e. the source of particles in units of $\rm cm^{3}s^{-1}${ ~\cite{baltzprop}. For the general case, 
\begin{align}
Q (\vec x, \epsilon) = \frac{1}{2} \frac{\rho^2(\vec x)}{m^2} \sigma v \frac{dN}{d\epsilon}
\end{align}
where $dN/d\epsilon$ is the spectrum of the resultant particles per annihilation as a function of energy. In the case of annihilation to electron and positron final states, $dN/d\epsilon = 2\delta(m)$, where $m$ is the mass of the dark matter particles. 
 In order to find a steady state solution for the spectrum $\partial n/d\epsilon$, the left hand side is set to zero and the solution is carried out as detailed in Ref~\cite{baltzprop}. 

We are interested in looking at the positron and electron spectrum and searching for dark matter signal. We assume an NFW dark matter halo profile ~\cite{nfw} with the local dark matter density, $\rho_0 = 0.43$ $\rm GeVcm^{-3}$. In particular, we consider a model where two dark matter particles of a particular mass annihilate into a positron and an electron, which then propagate through the galaxy and are detected on/near Earth. 

In this case, Q the source term depends on the dark matter annihilation cross section as well as the inhomogeneity of the distribution (leads to a boost factor which we take to be 1). The energy loss rate, $b = 10^{-16}(E/GeV)^2$, is the result of inverse Compton scattering on starlight and the cosmic microwave background, and synchrotron radiation due to the galactic magnetic field. Following Ref. ~\cite{hooperedge}, we expect a spectrum with an edge feature at the mass of the dark matter which is:

\begin{align}
\frac{dn}{d \epsilon} =  \frac{Q(m_{dm}, x_0)}{b}\theta(m_{dm}- \epsilon)
\end{align}

Following Ref.~\cite{baltzprop} we evaluate the size of this feature using $Q= n_0^2<\sigma v>m_{dm}^{-2} $ for a model dark matter that annihilates to electrons and positrons. Following Ref.~\cite{baltzprop} again, we  calculate that the detected flux at the edge of a particular dark matter mass  to be: 
\begin{align}
\frac{d\Phi}{d\epsilon}= \frac{c}{4\pi}\frac{dn}{d\epsilon}
\end{align}

For the case of $m_{dm}=$ 130 GeV,  with a cross section of $3$x$10^{-26}$ $\rm cm^{3}s^{-1}$, we calculated this edge size to be $E^3d\Phi/dE=$ $ 4.9 \rm\, GeV^2m^{-2}sr^{-1}s^{-1} $, which matches closely with the value of $4.8 \rm\,GeV^2m^{-2}sr^{-1}s^{-1}$ we obtained through our GALPROP simulation.

 To determine the cosmic ray spectrum as observed at the Solar System, we solve the standard propagation equation (using the publicly available code GALPROP ~\cite{Strong:1998pw}):
\begin{align} 
\frac{\partial   \psi }{\partial	 t } &=& Q(\textbf{r},p) + \bigtriangledown \cdot \left( D_{xx} \bigtriangledown \psi  - \textbf{V} \psi \right) +
\frac{\partial }{ \partial p  } p^2 D_{pp} \frac{\partial }{ \partial p  } \frac{1}{p^2} \psi   \nonumber\\
&& - \frac{\partial }{ \partial p  } \left[\dot{p} \psi -\frac{p}{3} \left( \bigtriangledown \cdot \textbf{V} \right) \psi \right]  
- \frac{1}{\tau_f} -\frac{1}{\tau_r} \psi   \ ,
\label{propeq}
\end{align}

where $\psi(\textbf{r},p,t)$ is the number density of a given cosmic ray species per unit momentum, and the source term $Q(\textbf{r},p)$ includes the products of the decay and spallation of nuclei, as well as any primary contributions from supernova remnants, pulsars, dark matter annihilations, etc. $D_{xx}$ is the spatial diffusion coefficient, which is parametrized by $D_{xx} = \beta D_{0xx} (\rho/4 GV)^\delta$, where $\beta$ and $\rho$ are the particle's velocity and rigidity, respectively. Also included in this equation are the effects of diffusive reacceleration and radioactive decay~\cite{Strong:1998pw}, however we neglect the effects of convection. The contribution to the source term, $Q(\textbf{r},p)$, from dark matter is simply determined by the flux of annihilation products injected into the halo. In our calculations, we adopt $D_{0xx}=4.02 \times 10^{28}$ cm$^2$/s and apply free-escape boundary conditions at 4 kpc above and below the Galactic Plane. These choices lead to boron-to-carbon and antiproton-to-proton ratios that are consistent with observations ~\cite{HooperSimet, simethooper}.

\begin{figure*}[!]
\vspace{-.5cm}
\includegraphics[width=0.46\textwidth ]{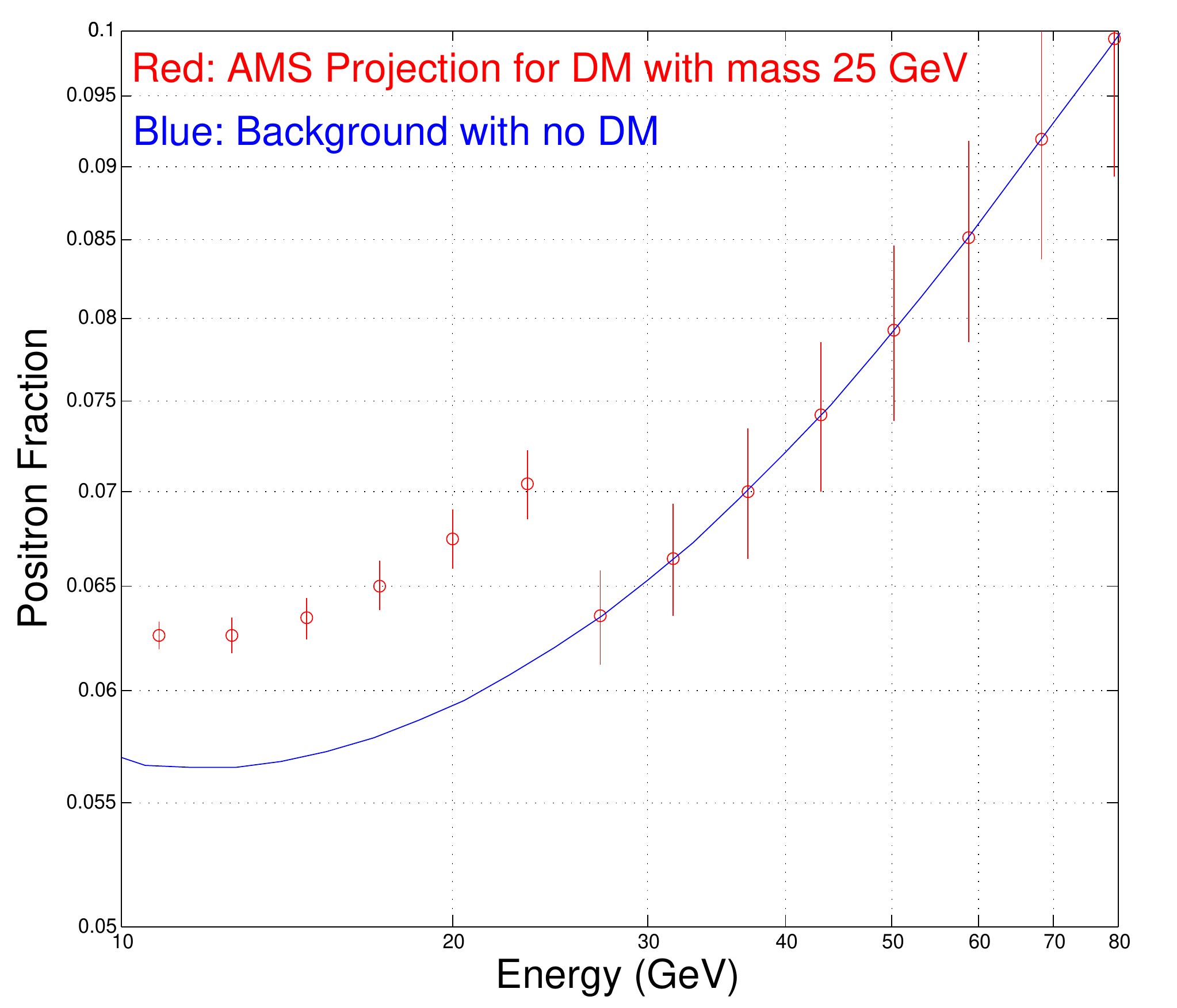}
\hspace{-1.0cm}
\hspace{1.0cm}
\vspace{.9cm}
\includegraphics[width=0.47\textwidth ]{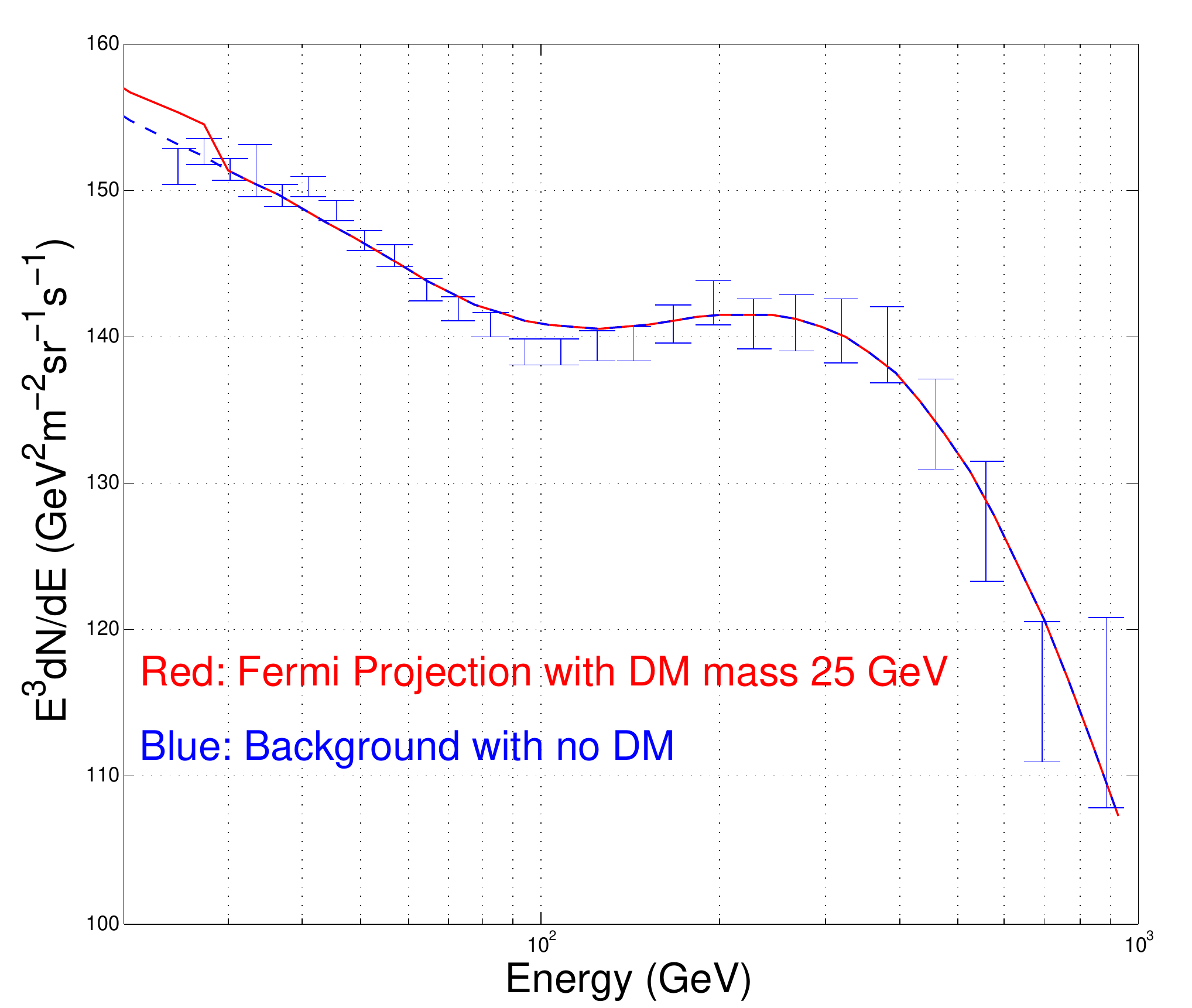}
\vspace{-1cm}

\vspace{0cm}
\includegraphics[width=0.46\textwidth ]{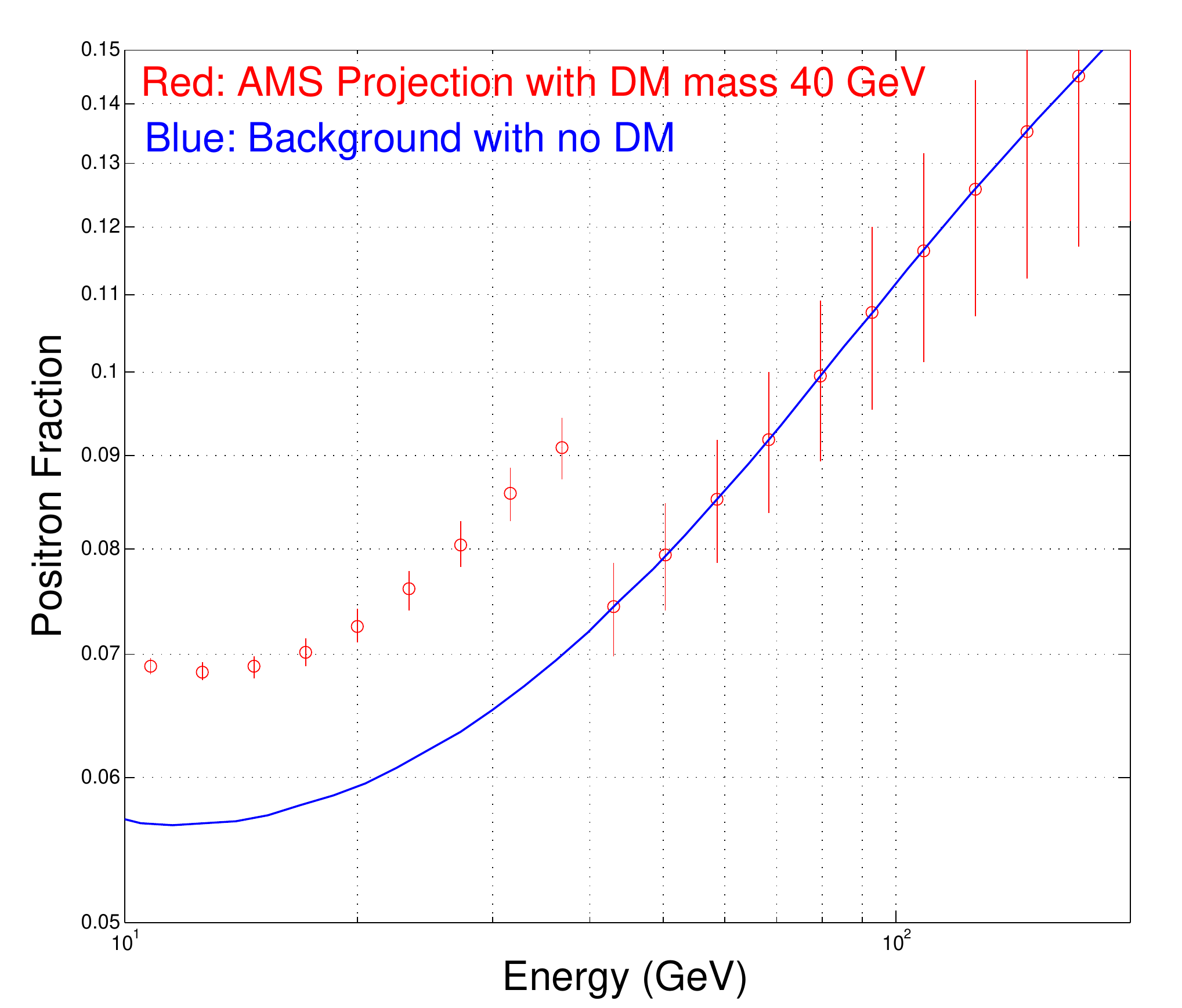}
\hspace{-1.0cm}
\hspace{1.0cm}
\vspace{.9cm}
\includegraphics[width=0.47\textwidth ]{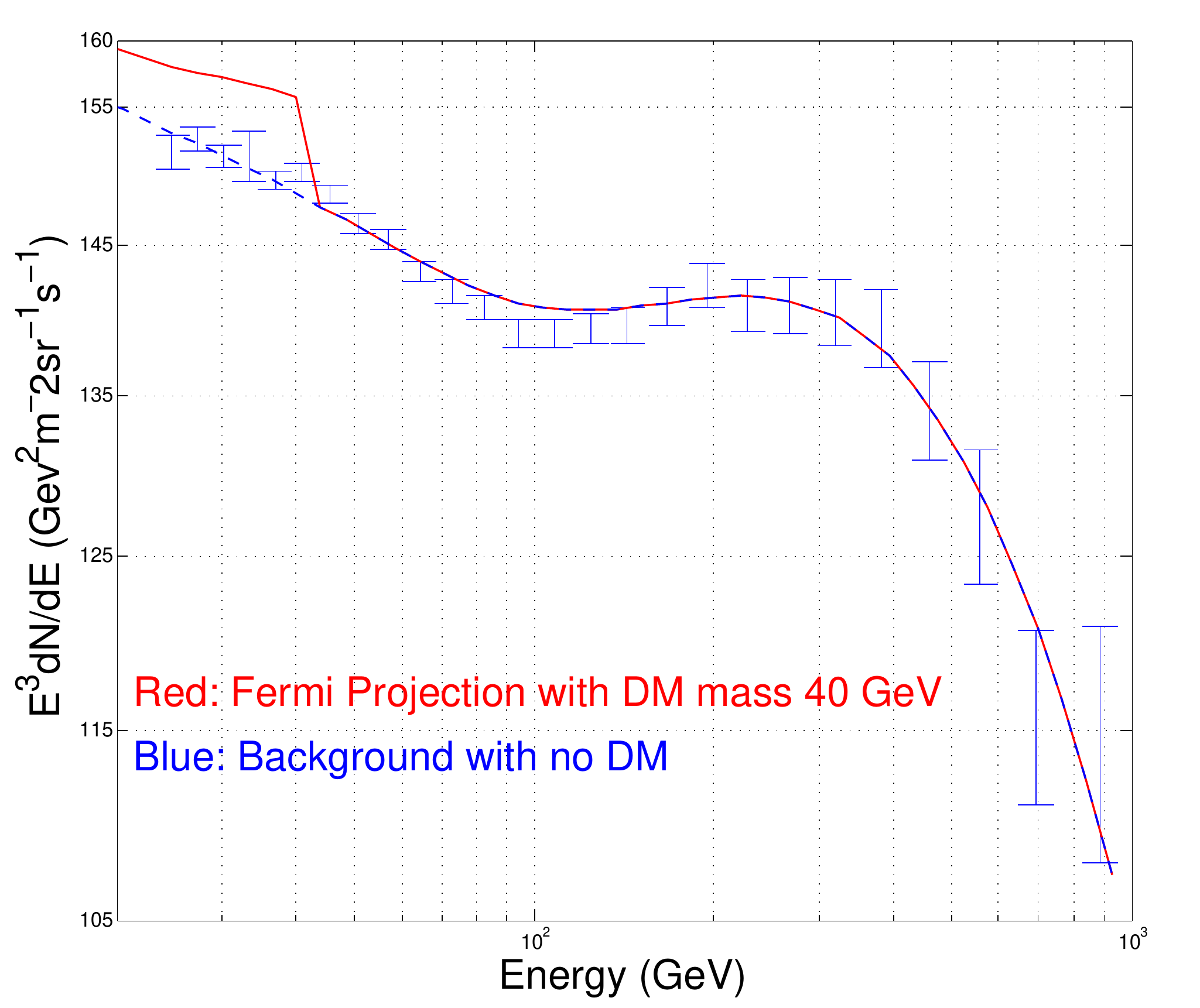}
\vspace{-1cm}

\vspace{0cm}
\includegraphics[width=0.46\textwidth ]{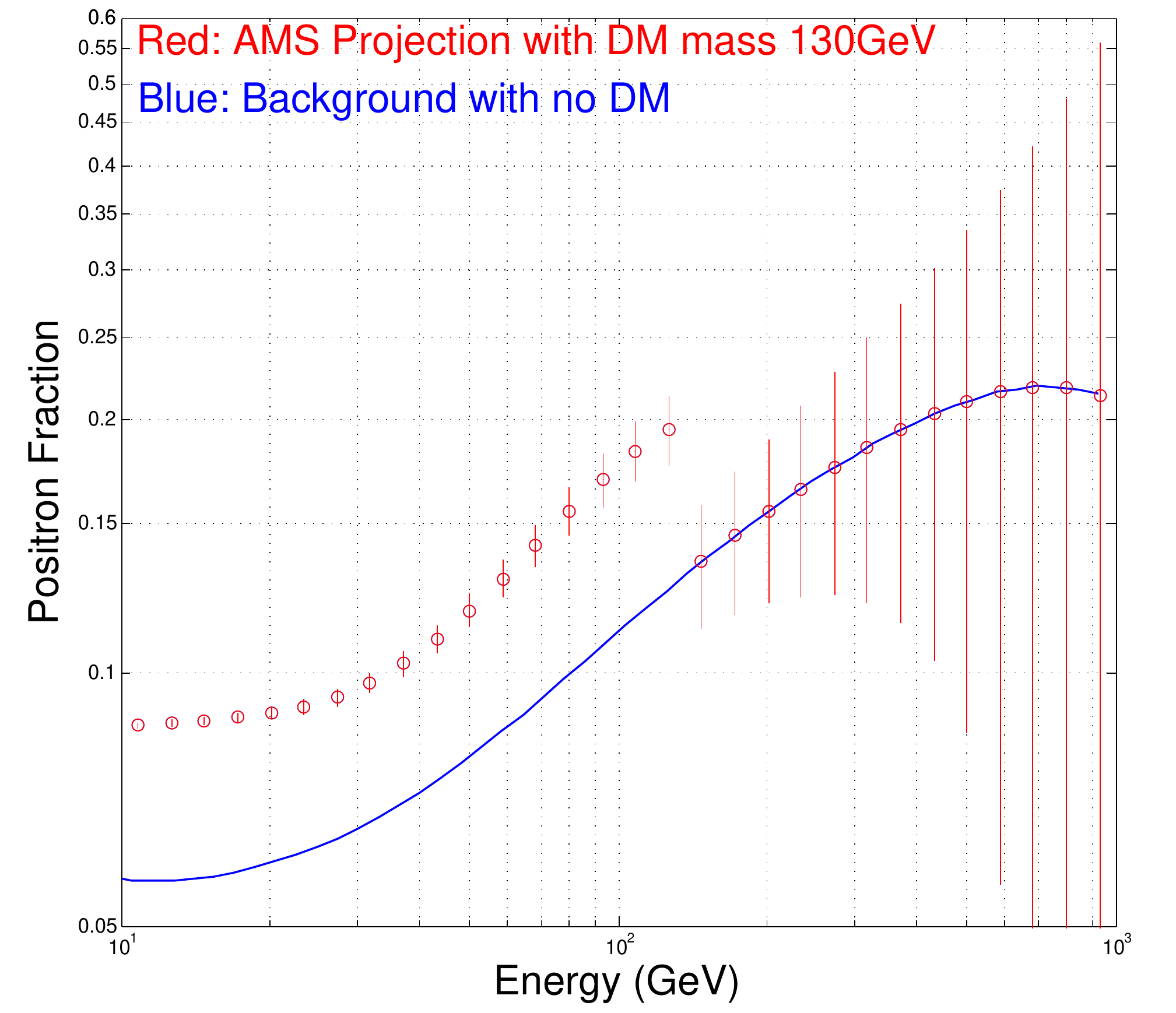}
\hspace{-1.0cm}
\hspace{1.0cm}
\vspace{.9cm}
\includegraphics[width=0.48\textwidth ]{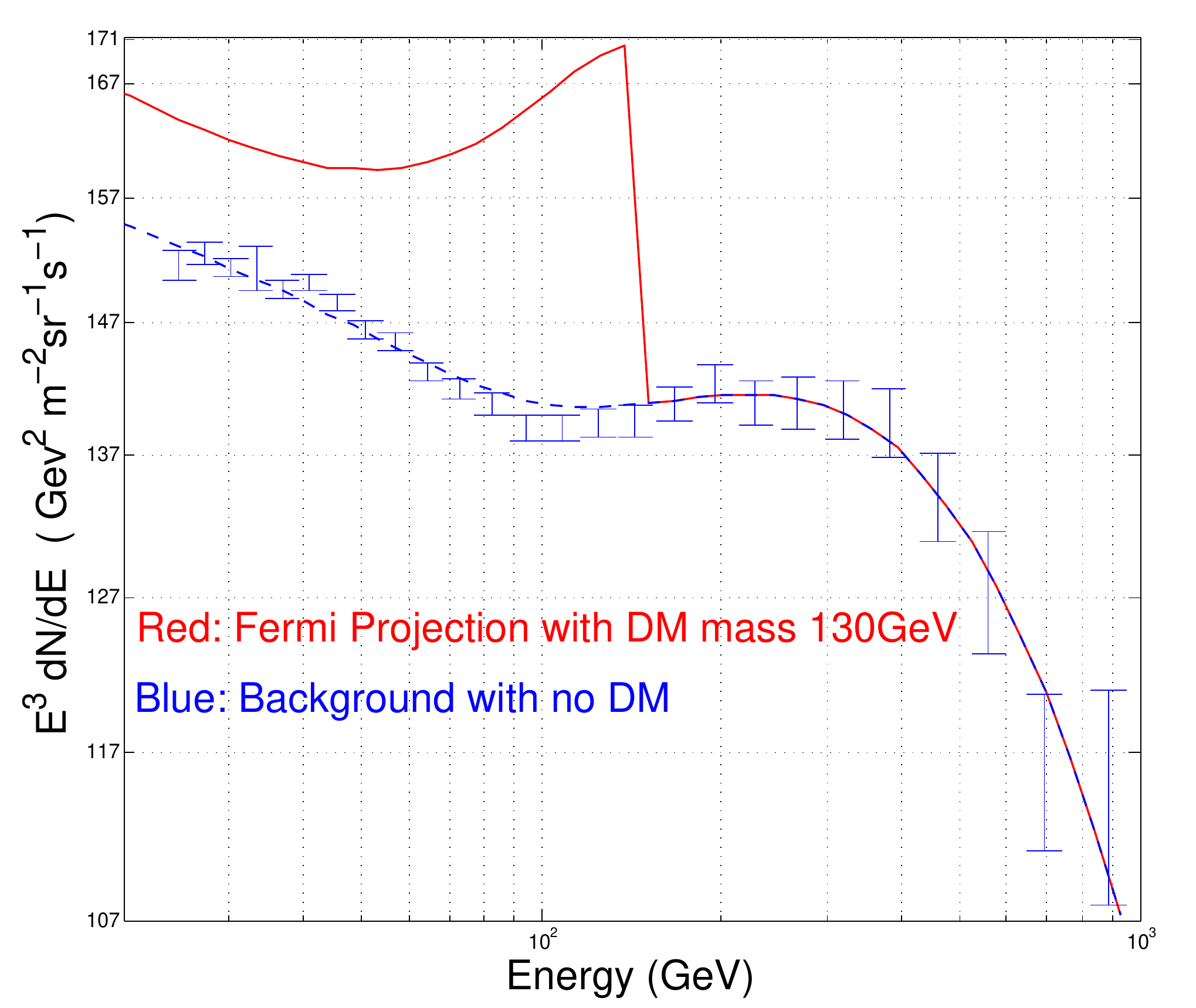}
\vspace{-1cm}
\caption{Left: Projection for AMS's measurement of the positron fraction for the case of dark matter with masses 25 GeV ($\sigma v= 1.5\times 10^{-27}$ cm$^3$/s), 40 GeV ($\sigma v= 1.5\times 10^{-26}$ cm$^3$/s) and 130 GeV ($\sigma v= 9\times 10^{-26}$ cm$^3$/s) annihilating to $e^+ e^-$. These cross sections were picked such that $\chi^2=4$ for the midpoint of the two bins adjacent to the step, corresponding to detection of a sudden spectral feature at the 95$\%$  confidence level. Right:  Total electron + positron flux, compared with Fermi data for the same dark matter scenarios shown in the left panels. Although we cannot rule out the possibility that AMS could detect a sudden spectral feature at energies below $\sim$40 GeV, it is clear that any higher energy feature potentially observable by AMS is already ruled out by Fermi. }
\label{DF2}
\end{figure*}

\begin{figure*}[!]
\vspace{-.5cm}
\includegraphics[width=0.48\textwidth ]{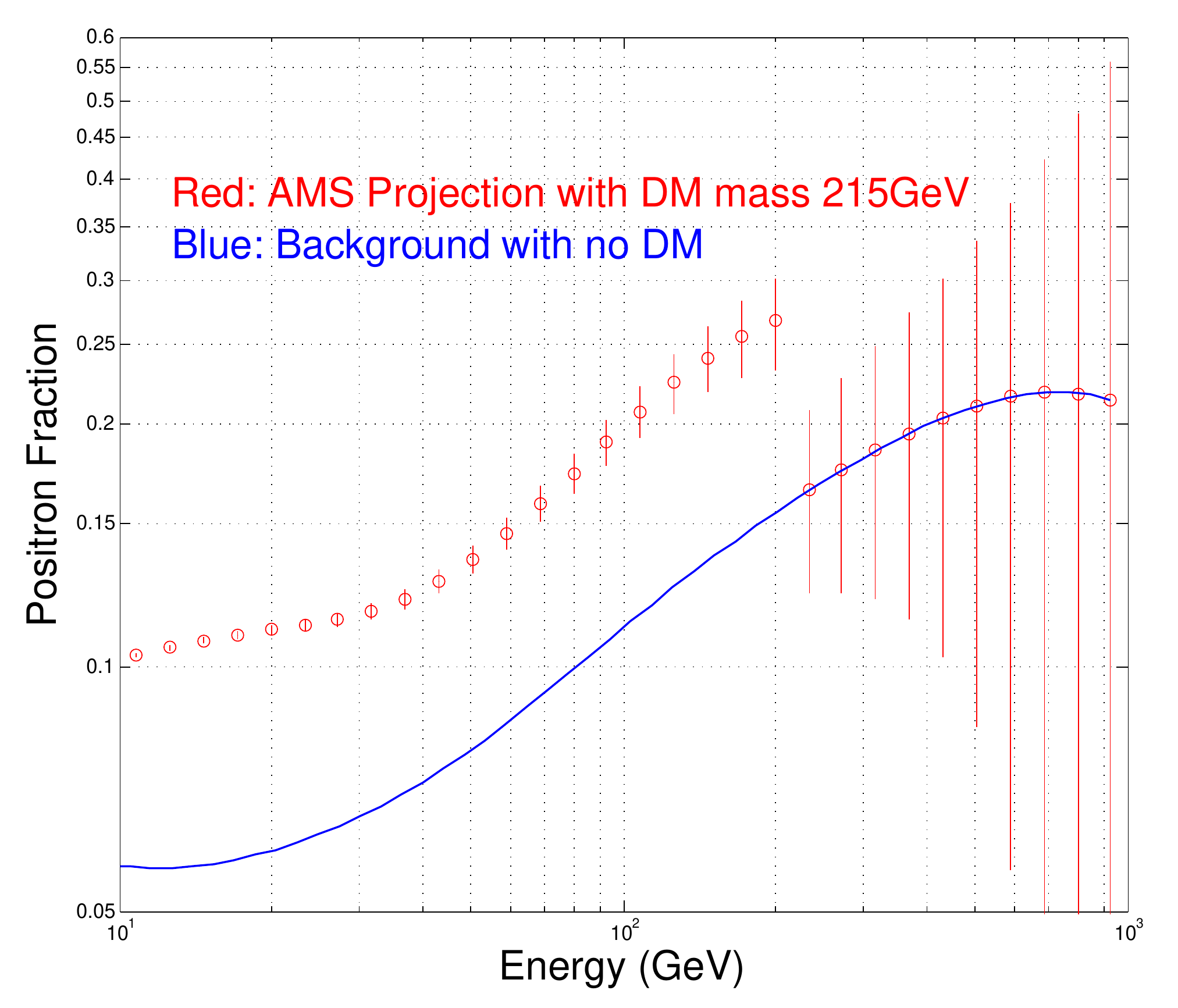}
\hspace{-1.0cm}
\hspace{1.0cm}
\vspace{.9cm}
\includegraphics[width=0.48\textwidth ]{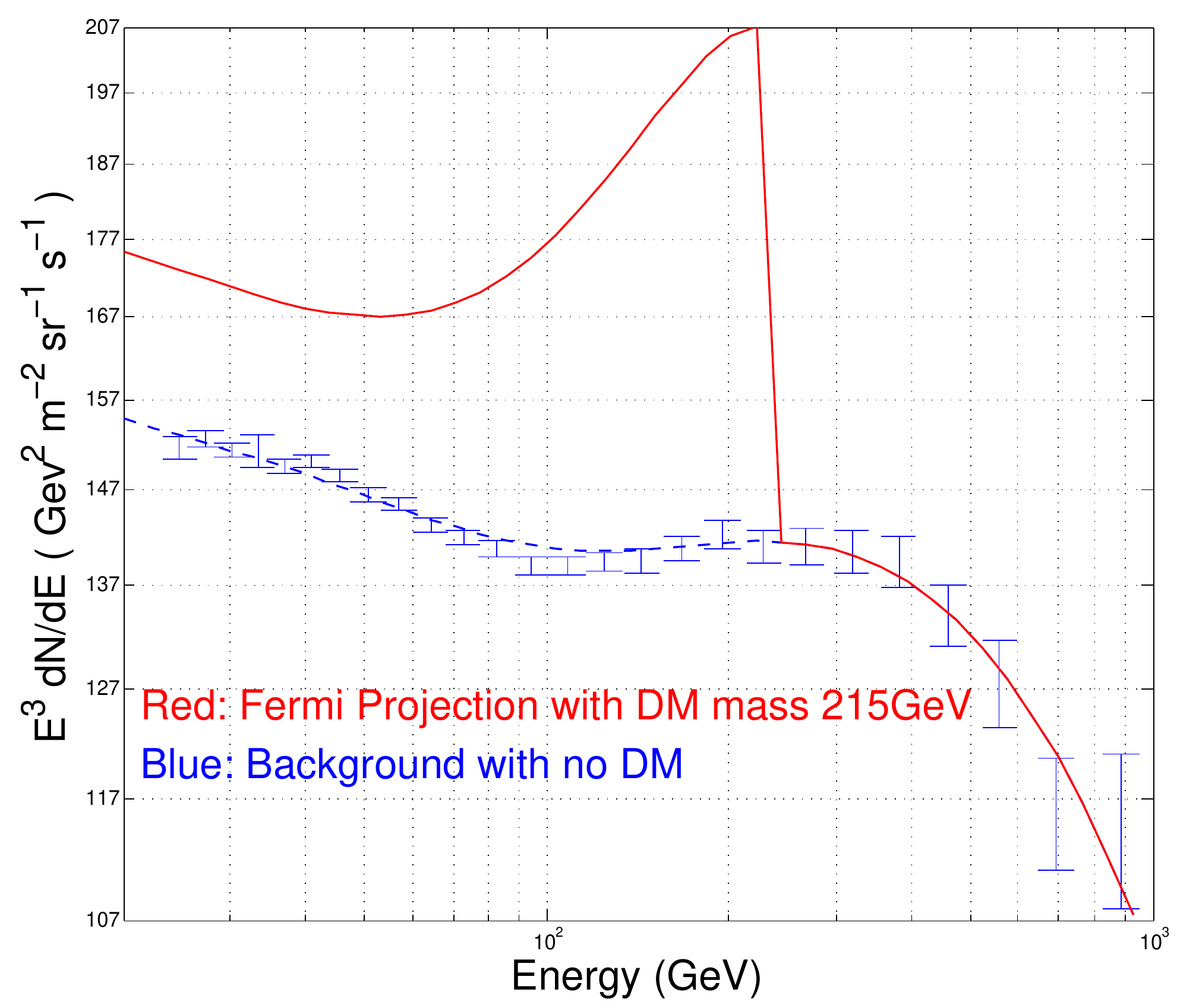}
\vspace{-1cm}

\vspace{--.5cm}
\includegraphics[width=0.48\textwidth ]{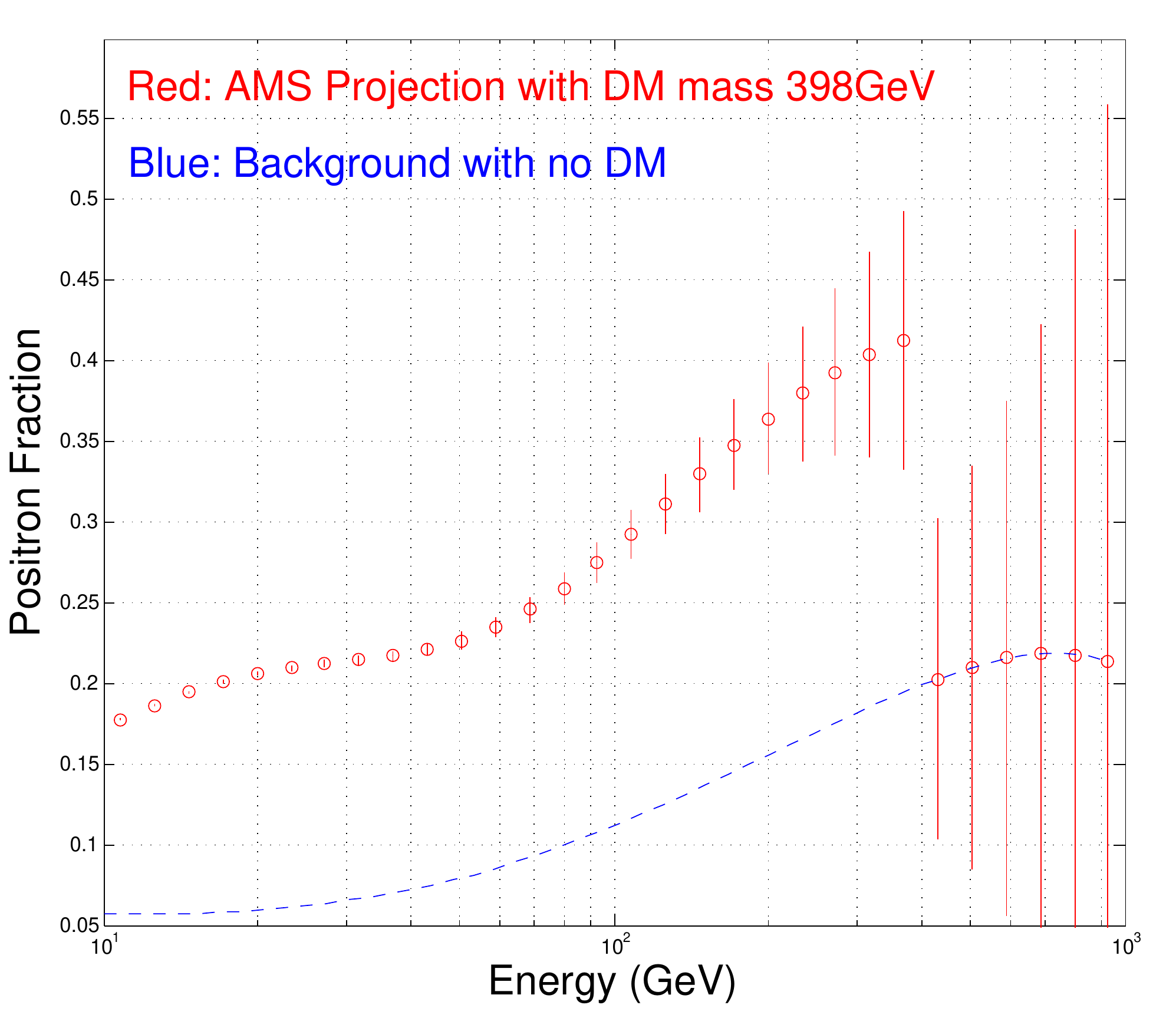}
\hspace{-1.0cm}
\hspace{1.0cm}
\vspace{.9cm}
\includegraphics[width=0.48\textwidth ]{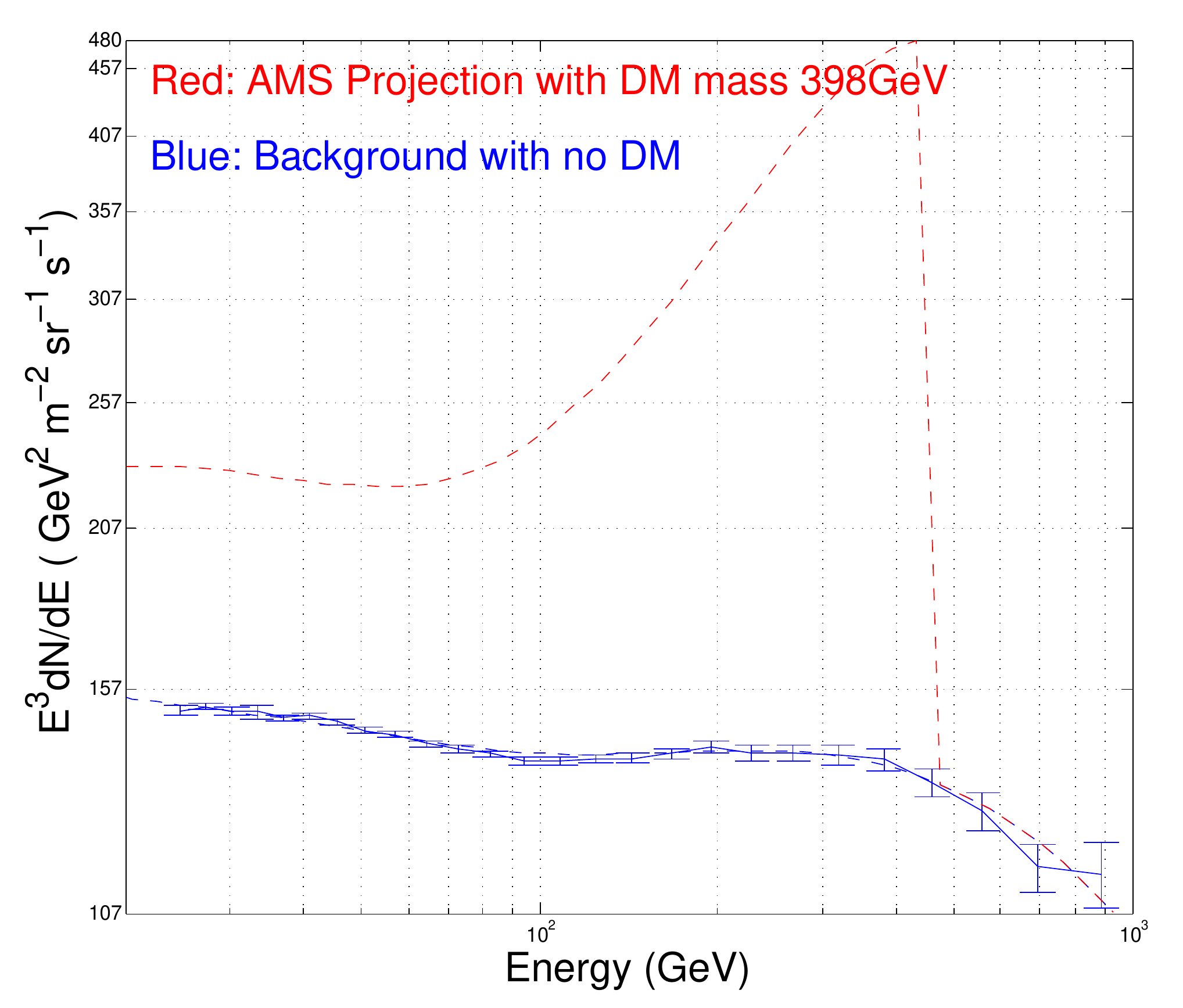}
\vspace{-1cm}
\caption{ Left: Projections for AMS measurement of positron fraction for the cases of dark matter with masses 215 ($\sigma v= 3.3\times 10^{-25}$ cm$^3$/s ) and 398 GeV($\sigma v= 3\times 10^{-24}$ cm$^3$/s) annihilating to $e^+, e^-$. These cross sections were picked such that $\chi^2=4$ for the midpoint of the two bins adjacent to the step, corresponding to detection of a sudden spectral feature at the 95$\%$  confidence level. Right:  Total electron + positron flux, compared with Fermi data. It is clear that these are ruled out by Fermi data.  Considering the trend, we conclude that the requisite cross section for higher masses is still greater and therefore we do not show further plots for higher masses.}
\label{DF3}
\end{figure*}

\section{ Astrophysical Sources of Cosmic Ray Electrons and Positrons}

We model the background positron and electron flux to match the measurements made by Fermi. Primary sources of electrons include supernova remnants and pulsars. Secondary electrons and positrons are those created by collisions of cosmic rays which occur during propagation through the galaxy. We account for the overall flux as a sum of these primary and secondary electrons and positrons combined with the flux from one nearby supernova remnant Monogem and the associated pulsar B0656+14. This pulsar is located 290 parsecs from the Solar System and is 110,000 years old~\cite{monogem}. We assume this pulsar to have injected a spectrum of positrons and electrons of the form, $Q  \propto E_e^{-1.7}$. We follow Ref.~\cite{Hooper:2008kg} in determining the flux of positrons and electrons at the Solar System from this pulsar and add this to the contribution predicted from primary and secondary production, as obtained using GALPROP. 

 The cosmic ray spectrum as observed by detectors close to the Earth is further affected by solar winds and heliospheric magnetic field~\cite{solarmod}.  This effect modeled by an effective potential, $\Phi = 0.4$ $ \rm GeV$, is especially important for energies smaller than roughly 20 GeV. The interstellar cosmic ray flux $ J_{IS}$ is related to the observed flux, $J$, as shown below: 
\begin{align}
J(p)= \frac{p(p+2m_p) }{(p+\Phi)(p+\Phi+ 2m_p) } J_{IS}
\end{align}

To project the error bars for AMS, we follow Ref.~\cite{Pato:2010im}. In particular, we convolve the spectrum of the positron fraction with an energy resolution of $\Delta E/E = \sqrt{(0.106/\sqrt{E({\rm GeV})})^2+(0.0125)^2}$ (corresponding to about 3.5\% at 10 GeV), an ability to reject protons from positrons and protons from electrons at the level of $3 \times 10^5$~\cite{Casaus:2009zz}, positrons from electrons at $1 \times 10^4$  and an acceptance of 0.045 m$^2$ sr. We are assuming 15 bins per decade. While we have calculated our error bars for 3 years of data taking, the systematic rather than statistical errors dominate the results.

In Figure ~\ref{leptonflux}, we show our background model. In the left frame of Fig.~\ref{leptonflux}, we show the result of our model along with Fermi data. In the right frame of Fig.~\ref{leptonflux}, we show the resulting positron fraction and compare this to that measured by PAMELA.

\section{Incompatibility with Fermi Constraints}
For a series of dark matter masses between 25 and 1000 GeV (which is the range covered by existing Fermi data), we determine the cross section required to produce a step in the positron fraction discernible by AMS. For a spectral feature that can be detected at the 95$\%$ confidence level,  we require $\chi^2\geq 4$ for the two bins adjacent to the step. 

When we plot the positron + electron spectrum alongside our fit to Fermi data, we find that in most cases the corresponding feature would be incompatible with the existing Fermi data. This is shown in Figures ~\ref{DF2} and ~\ref{DF3}. We can conclude that for all masses above $\sim$40 GeV, the existence of dark matter particles that annihilate into positrons and electrons to produce a feature detectable by AMS is already ruled out by Fermi. 
 One special point is the  the case of $\rm m_{dm}=$ 130 GeV. It has been speculated that a spectral gamma ray feature observed by Fermi around 130 GeV may be related to signals from dark matter annihilations to $\gamma \gamma$, $\gamma Z$ or $\gamma h$ ~\cite{130sp2, 130sp1}. As no continuum gamma ray signal is observed, however, there is a motivation to consider 130 GeV dark matter particles that annihilate to final states such as electrons and positrons ~\cite{hooper130} .  We show in row 3 of Figure ~\ref{DF2} the predictions for this dark matter mass. We conclude that  AMS is unlikely to observe a spectral feature associated with dark matter of this mass. 
\section{Conclusion}
Looking for dark matter is one of the missions of the AMS detector. We consider dark matter models where two dark matter particles annihilate to produce an electron-positron pair giving rise to a signature bump at the mass of the annihilating particles. From our calculations we see that for masses greater than $\sim$40 GeV, the cross sections corresponding to a detectable feature  in the positron fraction spectrum of AMS are already ruled out by existing Fermi measurements of the positron+ electron spectrum. 
\newline
\newline
Acknowledgements: 
We would like to thank  Dan Hooper, Wei Xue and Ilias Cholis for helpful discussions.







 \bibliographystyle{h-physrev}



\begin{thebibliography}{10}


\bibitem{FermiLAT:2011ab}
Fermi LAT Collaboration, M.~Ackermann {\em et~al.},
\newblock Phys.Rev.Lett. {\bf 108}, 011103 (2012), 1109.0521.


\bibitem{Adriani:2008zr}
PAMELA Collaboration, O.~Adriani {\em et~al.},
\newblock Nature {\bf 458}, 607 (2009), 0810.4995.

\bibitem{AMS}
AMS Collaboration (J. Alcaraz et al.). 
\newblock Physics Reports 366: 331Ð405, 2002. 74pp.


\bibitem{Baltz:2004ie}
E.~A. Baltz and D.~Hooper,
\newblock JCAP {\bf 0507}, 001 (2005), hep-ph/0411053.


\bibitem{hooperbertone}
G. Bertone, D. ~Hooper and J. Silk
\newblock Phys.Rept.405:279-390,2005, arXiv:hep-ph/0404175v2

\bibitem{hoopersilk}
D. ~Hooper and J. Silk
\newblock Phys.Rev. D71 (2005) 083503, arXiv:hep-ph/0409104

\bibitem{baltzprop}
E. A. Baltz and J. Edjso
\newblock Phys.Rev.D59:023511,1998

\bibitem{nfw}
J F. Navarro, C. S. Frenk and S. D. M. White
\newblock   Astrophys.J.462:563-575,1996


\bibitem{hooperedge}
E. A. Baltz and D. ~Hooper
\newblock JCAP 0507:001,2005


\bibitem{Strong:1998pw}
A.~Strong and I.~Moskalenko,
\newblock Astrophys.J. {\bf 509}, 212 (1998), astro-ph/9807150.



\bibitem{simethooper}
M. Simet and D. ~Hooper, 
\newblock JCAP 0908:003,2009


\bibitem{HooperSimet}
D. ~Hooper and K M ~Zurek, 
\newblock 	arXiv:0909.4163 [hep-ph]



\bibitem{monogem}
V. Barger et al. 
\newblock Phys.Lett.B678:283-292,2009.  0810.4994. arXiv:0904.2001 [hep-ph]



\bibitem{Hooper:2008kg}
D.~Hooper, P.~Blasi, and P.~D. Serpico,
\newblock JCAP {\bf 0901}, 025 (2009), 0810.1527.

\bibitem{solarmod}
L. J. Gleeson and W. I. Axford, Astrophys. J. 154 1011 (1968)




\bibitem{Pato:2010im}
M.~Pato, M.~Lattanzi, and G.~Bertone,
\newblock JCAP {\bf 1012}, 020 (2010), 1010.5236.


\bibitem{Casaus:2009zz}
J.~Casaus,
\newblock J.Phys.Conf.Ser. {\bf 171}, 012045 (2009).


\bibitem{130sp2}
C. Weniger
\newblock arXiv:1204.2797 [hep-ph]




\bibitem{130sp1}
D. P. Fiinkbeiner, M. Su, C Weniger
\newblock arXiv:1209.4562 [astro-ph.HE]



\bibitem {hooper130} 
M. R. Buckley and D. ~Hooper
\newblock  arXiv:1205.6811 [hep-ph]


\bibitem{Adriani:2008zq}
O.~Adriani {\em et~al.},
\newblock Phys.Rev.Lett. {\bf 102}, 051101 (2009), 





\bibitem{Profumo:2008ms}
S.~Profumo,
\newblock Central Eur.J.Phys. {\bf 10}, 1 (2011), 0812.4457.

\bibitem{Pato:2010ih}
M.~Pato, D.~Hooper, and M.~Simet,
\newblock JCAP {\bf 1006}, 022 (2010), 1002.3341.


\bibitem{Hooper:2011ti}
D.~Hooper and T.~Linden,
\newblock Phys.Rev. {\bf D84}, 123005 (2011), 1110.0006.





\bibitem{weixue}
D. ~Hooper and W. Xue, 
\newblock arXiv:1210.1220 [astro-ph.HE]

\bibitem{dgrasso}
D. Grasso et al. 
\newblock Astropart.Phys.32:140-151,2009



\bibitem{Baltzprop}
E A Baltz, J Edsjo, 
\newblock Phys.Rev.D59:023511,1998



\bibitem{Moskalenko:2001ya}
I.~V. Moskalenko, A.~W. Strong, J.~F. Ormes, and M.~S. Potgieter,
\newblock Astrophys.J. {\bf 565}, 280 (2002), astro-ph/0106567.





\end{thebibliography}

\end{document}